\documentclass[prb, reprint, longbibliography, amsmath]{revtex4-2}

\usepackage{graphicx}
\usepackage[colorlinks=true]{hyperref}
\usepackage{physics}
\hypersetup{
	citecolor = {blue},
	linkcolor = {blue},
	urlcolor = {blue}
}

\newcommand{\z}{\mathcal{Z}}
\renewcommand{\var}[1]{\mathrm{Var}\qty({#1})}

\begin{document}

\title{Theory for polaritonic quantum tunneling}

\author{Kalle S.~U.~Kansanen}
\email{kalle.kansanen@gmail.com}
\affiliation{Department of Physics and Nanoscience Center, University of Jyv\"askyl\"a, P.O. Box 35 (YFL), FI-40014 University of Jyv\"askyl\"a, Finland}

\date{October 20th, 2022}

\begin{abstract}
I investigate the tunneling decay rate of a polaritonic system formed by a strong coupling between a vacuum cavity mode and $N$ metastable systems.
Using a simple model potential, I find the instanton solutions controlling the low-temperature tunneling rate.
The resulting rate modification due to the cavity is proportional to the mean of the second power of the light-matter coupling. 
No collective effect that would enhance the rates by a factor of $\sqrt{N}$ is present, which is in line with the results in the thermal activation regime.
\end{abstract}

\maketitle

\section{Introduction}
Tunneling is a manifestation of quantum coherence:
quantum systems are able to surmount barriers they energetically should not due to their wave-like properties~\cite{ankerhold2007quantum}. 
The range of tunneling systems is broad: elementary particles in nuclear matter~\cite{balantekin1998quantum}, electrons in conductors~\cite{tersoff1983theory,beenakker2008colloquium}, magnetization in nanomagnets~\cite{sangregorio1997quantum}, and superconducting phase in superconducting circuits~\cite{voss1981macroscopic}.
Theoretically, tunneling can be understood as the quantum mechanical counterpart to classical thermal activation describing, for instance, chemical reactions~\cite{miller1975semiclassical,hanggi1990reaction,cao1996unified}.

Another type of quantum coherence can be seen when a coherent exchange of energy between two quantum mechanical systems happens.
A prime example of such coherent systems is polaritons which are the hybrid excitations of the vacuum electromagnetic field and molecular degrees of freedom.
Recently, it has been suggested that the formation of such coherent systems could affect chemistry which is still poorly understood~\cite{garcia2021manipulating,wang2021roadmap,hertzog2019strong,ribeiro2018polariton}.
In fact, a transition state theory calculation shows that all the polaritonic enhancements to the reaction rate scale as $1/N$ where $N$ is the number of molecules participating in the polariton~\cite{zhdanov2020vacuum,campos2020polaritonic,li2020origin}.
This is often attributed to the fact that the coupling to light induces only two energetically different polaritonic states, separated in energy by the Rabi splitting proportional to $\sqrt{N}$, while $N-1$ molecular states, the so-called dark states, remain energetically the same.

Motivated by the idea of polaritonic chemistry,
I focus on a related question whether there can be a genuine polaritonic quantum tunneling effect. 
This question arises naturally as the light-matter coupling changes the coherence properties of the system at hand.
It also induces collective behavior through the formation of polaritons.
In fact, the $N-1$ dark states are superpositions over the molecular states even though their energy does not change.

In this article, I present a model of $N$ metastable systems coupled to a cavity mode and investigate the effect of the common cavity mode on the low-temperature tunneling decay rate.
For a simple model potential, I analytically solve the polaritonic rate modification using path integral techniques in the semiclassical approximation.
Such solvable models are rare; there are only a few truly multidimensional problems in quantum tunneling that have been solved analytically~\cite{ankerhold2007quantum,rontani2012tunneling}.

In the low-temperature regime, the tunneling decay rate is dominated by \emph{instantons}.
I find the instanton solutions for the polaritonic system without friction.
As the main result, I find the polaritonic rate modification as a function of the number~$N$ of metastable systems. 
The tunneling decay rate is modified by a factor proportional to the single-molecule coupling constant and not by the Rabi splitting.
This shows that the cavity indeed induces a coherence effect but it is not a collective effect. Similar to the transition state theory calculation~\cite{zhdanov2020vacuum,campos2020polaritonic}, the polaritonic enhancements scale as $1/N$ if the Rabi splitting is fixed.
Therefore, the practical route to realizing the cavity-induced coherence is not in the collective strong coupling regime with large number of systems but rather in single systems with large couplings to the cavity.

\section{Semiclassical approximation to tunneling}
\begin{figure}
    \centering
    \includegraphics{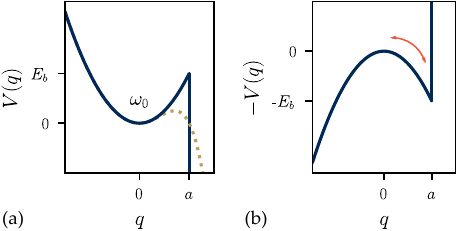}
    \caption{(a) Ski-jumping potential of Eq.~\eqref{eq:skijumppot}. It is obtained by a limiting process from a potential $E_b\qty[(q/a)^2 - \theta(q)(q/a)^n]$ with $\theta$ being the Heaviside step function and $n \rightarrow \infty$. The dotted line represents $n = 4$. (b) Inverted potential $-V(q)$. The arrow indicates the instanton solution in which the system moves from $q = 0$ to $q = a$ and back.}
    \label{fig:potential}
\end{figure}

Consider a metastable system described by a potential
\begin{align}
    V(q) = 
    \begin{cases}
        \frac{1}{2} \omega_0^2 q^2, & q \leq a,  \\
        - \infty, & q > a,
    \end{cases}
    \label{eq:skijumppot}
\end{align}
where $a$ determines the energy of the potential barrier $E_b = \frac{1}{2}\omega_0^2 a^2$ as in Fig.~\ref{fig:potential}(a).
The quadrature $q$ is defined here so that the conjugate momentum quadrature $p$ is given simply by $p = \dot q$.
Although this potential has been used before~\cite{grabert1984quantum,altlandsimons}, it lacks a name and, so, I call it the ski-jumping potential.
I set $\hbar = 1$ everywhere.

Next, consider $N$ identical metastable systems coupled to a single harmonic cavity mode whose position quadrature is $x$, normalized similarly to $q$.
I assume that this coupling is directly between the quadratures $x$ and $q$.
The total Hamiltonian of this polaritonic system is given by
\begin{align}
    H = \frac{1}{2} {\dot x}^2 + \sum_{i=1}^N \frac{1}{2} {\dot q}_i^2 + V_\mathrm{tot},
\end{align}
where
\begin{align}
    V_\mathrm{tot} = \sum_{i=1}^N V(q_i) + \frac{1}{2} \omega_c^2 x^2 +  \sum_{i=1}^N \lambda_i^2 x q_i.
\end{align}
Here, the apparent eigenfrequency of the cavity mode is $\omega_c$ while the light-matter coupling is encoded within $\lambda_i^2$. 
It can be related to the coupling constant $g_i$ obtained from a quantum electrodynamics calculation~\cite{walls2007quantum} by $\lambda_i^2 = \sqrt{\omega_c \omega_0} g_i$.
If one considers the ski-jumping potential to be a simplistic model of a potential energy surface of a molecule, the exact value of each coupling constant depends on the orientation and position of the molecule within the cavity~\cite{walls2007quantum}.

There are several methods to calculate the tunneling decay rate of a metastable system.
Here, I use the so-called $\Im F$ method~\cite{langer1967theory} as it can straightforwardly be used for multidimensional systems and provides the possibility to extend the theory to include dissipation~\cite{ankerhold2007quantum,altlandsimons}.
Physically, the idea is simple: the metastability of the ski-jumping potential means that there are no stationary states.
This may be represented by the eigenenergies obtaining a finite imaginary part which is associated to a tunneling decay rate.
Likewise, the partition function~$\z$ defined in terms of the states within the ski-jumping potential obtains an imaginary part.
Then, the tunneling decay rate $k$ at low temperature may be expressed as
\begin{align}
    k = \frac{2}{\beta} \Im \ln \z,
\end{align}
where $\beta = 1/k_B T$ is the inverse temperature~\cite{ankerhold2007quantum,kleinertbook}.
The partition function~$\z$ can be represented by an Euclidean path integral
\begin{align}
    \z = \int D(\phi) \exp{-S_E[\phi(\tau)]}
\end{align}
over $\beta$-periodic paths in imaginary time $\tau = it$.
Here, $\phi = (x, q_1, \dots, q_N)^T$ represents a column vector of all the dynamical degrees of freedom and the Euclidean action is given by
\begin{align}
    S_E = \int_{-\beta/2}^{\beta/2} \dd{\tau} \qty[\frac{1}{2}\dot{\phi}^T \dot{\phi} + V_\mathrm{tot}(\phi)].
\end{align}
One can associate this Euclidean action to the classical action of systems moving in the inverted potential $-V_\mathrm{tot}$.

For independent systems, the total potential energy can be written as a sum of system's potential energies.
Thus, the partition function factorizes as $\z = \z_1^N$ for identical systems.
Whatever the single-system tunneling decay rate $k_1$ is, the total rate is then $k = N k_1$.

In general, solving the path integral exactly to obtain the partition function is difficult.
Thus, I resort to the semiclassical approximation which is valid when the barrier energy $E_b$ is large compared to the real part of the ground state energy (which is of the order of $\omega_0$)~\cite{kleinertbook}.
I expand the path integral around the classical solutions and take into account only the quadratic fluctuations
\begin{subequations}
\begin{align}
    \z &\approx \sum_\mu I_\mu e^{-S_E(\phi_\mu)} , \\
    I_\mu &= \int D(r_\mu) \exp{- \frac{1}{2}r_\mu^T\qty[\partial_\tau^2 + \mathcal{V}(\phi_\mu)]r_\mu}. \label{eq:gaussianfluctuations}
\end{align}
\end{subequations}
Here, $\phi_\mu$ represents one possible classical $\beta$-periodic path and $I_\mu$ the contribution of quadratic fluctuations which may be expressed using a second derivative matrix $\mathcal{V}_{ij} = \partial^2 V_\mathrm{tot}/\partial \phi_i \partial \phi_j$ evaluated at the corresponding classical solution $\phi_\mu$. 
The integration variable $r_\mu$ is the deviation from $\phi_\mu$ with the boundary conditions $r_\mu(\pm\beta/2) = 0$.
As the action $S_E$ is a real variable, the imaginary part of the partition function must be in fluctuations $I_\mu$.

The ski-jumping potential allows for the solution of classical paths in a general case but it complicates the evaluation of the fluctuations as the potential is discontinuous at $q_i = a$.
These problems can mostly be avoided since the quadratic fluctuations can be expressed in terms of the classical solutions exactly in the case of a closed system~\cite{dashen1974nonperturbative,liang1992bounces}.

\subsection{Solution of the Euclidean action}
First, I solve the classical periodic paths in imaginary time.
The problem is the same as solving classical motion in real time but in the inverted potential. 
Note that $q_i = a$ represents a wall in the inverted potential as in Fig.~\ref{fig:potential}(b).
Thus, at this point, the velocity~$\dot{q}_i$ is discontinuous.
Rather than trying to piece together solutions before and after hitting the wall, I expand the mathematical trick presented in Ref.~\onlinecite{grabert1984quantum} for a polaritonic system and take this discontinuity into account at the level of the equations of motion.
If a single quadrature~$q_1$ hits the wall at time~$\tau_1$, that is, $q_1(\tau = \tau_1) = a$,
the dynamics in the inverted potential is determined by
\begin{subequations}\label{eq:EOMS}
\begin{align}
    - \ddot{x} + \omega_c^2 x +  \sum_{i=1}^N \lambda_i^2 q_i &= 0, \\
    - \ddot{q}_1 + \omega_0^2 q_1 + \lambda_1^2 x &= A \delta(\tau - \tau_1),\\
    - \ddot{q}_i + \omega_0^2 q_i + \lambda_i^2 x &= 0, \qq{$i = 2,3,\dots N$.}
\end{align}
\end{subequations}
The unknown constant~$A$ is determined from the condition $q_1(\tau = \tau_1) = a$.

Since I am searching for periodic solutions, the way to proceed is to write all dynamical quantities as Fourier series.
Here, I choose the convention $f(\tau) = \sum_m f_m e^{i \omega_m \tau}$ with $\omega_m = 2 \pi m/\beta$ being the bosonic Matsubara frequency.
The inverse transformation is then $f_m = \frac{1}{\beta}\int\dd{\tau} f(\tau)e^{-i\omega_m\tau}$.
By applying the latter definition to Eqs.~\eqref{eq:EOMS}, I find
\begin{subequations}
\begin{align}
    (\omega_c^2 + \omega_m^2) x_m + \sum_{i=1}^N\lambda_i^2 q_{i,m} &= 0, \\
    (\omega_0^2 + \omega_m^2) q_{1,m} + \lambda_1^2 x_m &= \frac{A}{\beta} e^{- i \omega_m \tau_1}, \label{eq:solstep1b}\\
    (\omega_0^2 + \omega_m^2) q_{i,m} + \lambda_i^2 x_m &= 0.\label{eq:solstep1c}
\end{align}
\end{subequations}
This set of linear equations can be solved.
The idea is first to find the dynamics of the cavity mode~$x$ which then gives the solutions of the individual quadratures~$q_i$. 
This is achieved by defining a collective variable $Q_m = \sum_{i=1}^N \frac{\lambda_i^2}{\expval{\lambda^2}}q_{i,m}$ with $\expval{\lambda^2} = \sum_i \lambda_i^2/N$ representing the average over the couplings.
The dynamics of $Q$ can be determined from Eqs.~\eqref{eq:solstep1b}--\eqref{eq:solstep1c}, which allows for solving the dynamics of $x$.
After a short calculation I find the solutions in Fourier space to be
\begin{subequations}
\begin{align}
    x_m &= - \frac{A}{\beta}\lambda_1^2 \chi_P(\omega_m) e^{-i \omega_m \tau_1}, \\
    q_{i,m} &= \frac{A}{\beta} \frac{\lambda_1^2\lambda_i^2}{\omega_m^2 + \omega_0^2} \chi_P(\omega_m) e^{-i \omega_m \tau_1}, \\
    q_{1,m} &= \frac{A}{\beta} \frac{1}{\omega_m^2 + \omega_0^2}\qty[1 + \lambda_1^4\chi_P(\omega_m)]e^{-i \omega_m \tau_1},
\end{align}
\end{subequations}
where I defined a short-hand notation describing the polaritonic response
\begin{align}
    \chi_P(\omega_m) = \qty[(\omega_m^2 + \omega_0^2)(\omega_m^2 + \omega_c^2) - N \expval{\lambda^4}]^{-1}.
    \label{eq:polresponse}
\end{align}
The cavity-mediated interaction can be seen in the fact that the dynamics of all the quadratures~$q_i$ depend on the coupling~$\lambda_1^2$ of the first quadrature.

The abstract Fourier space solutions become clearer in the zero-temperature limit $\beta \rightarrow \infty$.
Then, the Fourier series can be transformed to an integral which I evaluate using the residue theorem.
Setting $\tau_1 = 0$ for brevity, this results in the imaginary-time paths
\begin{subequations}\label{eq:instpath}
\begin{align}
    x(\tau) &=  A \frac{\lambda_1^2}{\sqrt{\expval{\lambda^4}}} \sqrt{\frac{1 - \delta^2}{4N}} \qty( \frac{e^{-\omega_+ \abs{\tau}}}{\omega_+} - \frac{e^{-\omega_- \abs{\tau}}}{\omega_-}), \\
    q_i(\tau) &= A \frac{\lambda_1^2\lambda_i^2}{\expval{\lambda^4}} \frac{f(\tau)}{N}, \\
    q_1(\tau) &= A \frac{e^{-\omega_0 \abs{\tau}}}{2 \omega_0} + A \frac{\lambda_1^4}{\expval{\lambda^4}} \frac{f(\tau)}{N}, \label{eq:inst1}\\
    f(\tau) &= \frac{1+\delta}{2}\frac{e^{-\omega_+ \abs{\tau}}}{2\omega_+} + \frac{1-\delta}{2}\frac{e^{-\omega_- \abs{\tau}}}{2\omega_-} - \frac{e^{-\omega_0 \abs{\tau}}}{2 \omega_0}
\end{align}
\end{subequations}
with further definitions of the polariton eigenfrequencies~$\omega_\pm$ without the rotating wave approximation and a detuning parameter~$\delta \in [-1,1]$ given by
\begin{subequations}
\begin{align}
    \omega_\pm &= \sqrt{\frac{\omega_0^2 + \omega_c^2}{2} \pm \frac{1}{2}\sqrt{4N \expval{\lambda^4} + (\omega_0^2 - \omega_c^2)^2}}, \\
    \delta &= \frac{\omega_0^2 - \omega_c^2}{\omega_+^2 - \omega_-^2}.
\end{align}
\end{subequations}
The Rabi splitting is typically defined as $\omega_+ - \omega_-$ when the cavity is on resonance $\omega_c = \omega_0$.
Finally, $A$ resolves by demanding that $q_1(\tau = \tau_1 = 0) = a$.
It gives rise to a weighted harmonic average
\begin{align}
    A = 2 a \frac{N \expval{\lambda^4}}{\frac{N \expval{\lambda^4} - \lambda_1^4}{\omega_0} 
    + \lambda_1^4\qty(\frac{1+\delta}{2}\frac{1}{\omega_+} + \frac{1-\delta}{2}\frac{1}{\omega_-})} \equiv 2 a \omega_{H,1}.
    \label{eq:hf}
\end{align}
Here, the weights are the second-order coupling constants $\lambda_i^4$ and detuning factors $(1 \pm \delta)/2$.
This expression already shows that, similarly to the discussion about dark states, the bare frequencies~$\omega_0$ are weighted with a factor proportional to $N - 1$ whereas the polariton frequencies~$\omega_\pm$ have a weight close to unity, independently of $N$.
Thus, in general, $\omega_{H,1} \approx \omega_0$ for $N \gg 1$.
If $\lambda_1^2 = 0$, then $\omega_{H,1} = \omega_0$.

\begin{figure}
    \centering
    \includegraphics{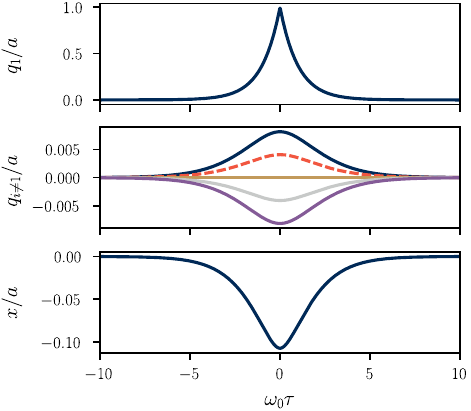}
    \caption{Polaritonic instanton solution for $N = 6$ on resonance $\omega_c = \omega_0$. The coupling constants are chosen so that $\lambda_1^2/\omega_0^2 = 0.1$ and $\lambda_{i\neq 1}^2/\omega_0^2 \in \qty{0, \pm 0.1, \pm 0.2}$. The dashed orange line in the middle graph also represents the second term in Eq.~\eqref{eq:inst1} that describes the modification to the bounce due to the light-matter coupling.}
    \label{fig:inst}
\end{figure}

An example of the polaritonic instanton solution is shown in Fig.~\ref{fig:inst}.
Initially, $\tau \rightarrow -\infty$, all the quadratures are at zero.
Very slowly, the first quadrature starts to evolve, pulling all the other systems with it.
The exact direction the other systems are pulled towards depends on the relative signs of the coupling constants~$\lambda_i^2$.
At time $\tau = 0$, the first quadrature is at the wall and bounces back.
From this hitting time to $\tau \rightarrow \infty$, the inverse happens.
The first quadrature starts to slow down and all the quadratures creep towards their initial position.

The Euclidean action follows directly from the instanton solutions at any temperature.
I obtain
\begin{subequations} \label{eq:action:result}
\begin{align}
    S_{E,1} &= \frac{1}{2} a^2 \qty[\frac{1}{\beta}\sum_m \frac{1 + \lambda_1^4 \chi_P(\omega_m)}{\omega_0^2 + \omega_m^2}]^{-1}\\
    &\rightarrow 2 \frac{E_b}{\omega_0} \frac{\omega_{H,1}}{\omega_0} \equiv S_0 \frac{\omega_{H,1}}{\omega_0}, \qq{when $\beta \rightarrow \infty$.} 
\end{align}
\end{subequations}
In the low-temperature limit, the action is determined by two ratios: 
First, the barrier energy~$E_b$ is compared to the pseudo-eigenenergy~$\omega_0$.
This is in contrast to the high-temperature result with $S_E = \beta E_b$.
Second, the polaritonic effect is contained within the ratio of the harmonic mean frequency~$\omega_{H,1}$ and the bare frequency~$\omega_0$.
This ratio is unity when there is no coupling, $\lambda_1^2 = 0$, and the action is just the bare action, $S_{E,1} = S_0$.
A fully uncoupled system does not know about the polaritons, as expected.

The classical action in Eq.~\eqref{eq:action:result} assumes that a single quadrature bounces off the wall or, more technically, $q_1(\tau = \tau_1) = a$ while $q_{i\neq1}(\tau) \neq a$ for all $\tau$.
This assumption is readily lifted and one can search for ``multi-bounce'' solutions with $M$ bounces, that is, $q_i(\tau_i) = a$ for $i = {1,2, \dots, M}$ and $M \leq N$.
A possible approach to obtain these classical solutions is briefly discussed in Appendix~\ref{app:multibounce}.

However, the multi-bounce solutions do not contribute to the quantum tunneling rate,
as they represent extremal points of the action~$S_E$ which are not saddle points.
This is an aspect of the quantum tunneling theory~\cite{coleman1988quantum} and the classical transition state theory~(TST)~\cite{hanggi1990reaction,hanggi1986escape} that is present when the metastable system has multiple degrees of freedom.
In the classical TST --- which is the high-temperature limit of the $\Im F$ approach used here --- the action is expanded around solutions that are constant in imaginary time.
In this case, the saddle points of the action~$S_E$ are directly the saddle points of the total potential~$V_\mathrm{tot}$.
These points are characterized by a single unstable quadrature~\cite{tstnote}.
I note that some works appear to be in contradiction with this principle, for instance Ref.~\onlinecite{yang2021quantum} suggests a ``coherent TST picture'' in the context of polaritonics.
Here, it is difficult to show that the multi-bounce instantons are not proper saddle points of $S_E$ because it would require evaluating the eigenvalue spectrum of the operator $\partial_\tau^2 + \mathcal{V}(\phi)$ defined in Eq.~\eqref{eq:gaussianfluctuations}.
However, there are two clear physical signs why the multi-bounce solutions must be neglected.
First, the high-temperature limit of those solutions correspond to saddle points with multiple unstable quadratures which I show in a special case in the Appendix~\ref{app:multibounce}.
Second, one does not obtain the correct limit of $N$ independent systems when the light-matter coupling vanishes, that is, $\lambda_i^2 \rightarrow 0$.
The latter point becomes more clear in the next section as I obtain the correct limit with single bounces.

\subsection{Polaritonic tunneling rate modification}
To get from the instanton solution~\eqref{eq:instpath} to the polaritonic tunneling decay rate, one needs to calculate the fluctuation factor~$I_\mu$.
The program is somewhat cumbersome even in the one-dimensional case~\cite{kleinertbook}. 
The first derivative of the instanton solution happens to be a zero eigenvalue mode for the fluctations and $I_\mu$ formally diverges.
The existence of the zero mode also implies that there exists a negative eigenvalue mode which makes $I_\mu$ imaginary.
A further complication is that one should include multiple sequential bounces.
Such paths are obtained by essentially glueing instanton solutions together:
The imaginary-time axis can be separated into $n$~partitions of length~$\beta/n$.
Since the instanton paths change appreciably only for the imaginary time~$2/\omega_0$, using the instanton solution~\eqref{eq:instpath} for each partition of length~$\beta/n$ gives a path with $n$~bounces.
The error of this process is exponentially small in $\beta$ when $2/\omega_0 \ll \beta$.
In these steps, I follow closely the one-dimensional treatment of Ref.~\onlinecite{liang1992bounces}.
I do this as the relatively recent literature~\cite{milnikov2002decay,erakovic2020instanton} 
cannot be applied since the instanton solution~\eqref{eq:instpath} is not differentiable at the hitting time.

The fluctuation factor of a single bounce can be obtained from a version of the Gelfand--Yaglom formula \cite{kleinertbook}
\begin{align}
    I^{n=1}_1 \propto - i \beta \sqrt{S_{E,1}} \sqrt{\frac{\epsilon_1(\beta)}{D_1}},
    \label{eq:onebounce:fluc}
\end{align}
where $D_1 = \abs{\det(\pdv{\phi_j(\beta/2)}{\dot \phi_i(-\beta/2)})}$ is the fluctuation determinant evaluated in the $\beta \rightarrow \infty$ limit and $\epsilon_1(\beta)$ provides a finite temperature correction to it.
In these and the following expressions, I denote the number of bounces as a superscript whereas the subscript refers to the quadrature that hits the wall.
I also choose not to keep track of the powers of $2\pi$; in the end, they are fixed by comparing to the non-interacting result.
By expanding the method in Ref.~\onlinecite{gildener1977pseudoparticle} to the multidimensional system at hand, I find
\begin{equation}
    \epsilon_1(\beta) \approx 2\frac{\dot \phi^T(-\beta/2) \ddot \phi(-\beta/2) - \dot \phi^T(\beta/2) \ddot \phi(\beta/2)}{\int_{-\infty}^\infty \dot\phi^T(\tau) \dot \phi(\tau) \dd{\tau}},
    \label{eq:finiteTcorr}
\end{equation}
where $\phi$ refers to the vectorized form of the instanton solution~\eqref{eq:instpath}.
The derivation of this result can be found in Appendix~\ref{app:ftc}.
The factor $-i$ is the Maslov--Morse index, which takes into account the one negative eigenvalue mode.
Mathematically, it follows from the singularity of the fluctuation determinant at the turning point of the classical solution (see e.g. Ref.~\onlinecite{kleinertbook}).
Lastly, $\beta \sqrt{S_{E,1}}$ follows from the Faddeev--Popov method as the hitting time~$\tau_1$ is in fact a free parameter.
By a change of integration variables from the zero mode proportional to the first derivative of the instanton solution to $\tau_1$ in $I_\mu$, one integrates $\tau_1$ over the whole range~$[-\beta/2, \beta/2]$ while the Jacobian of the transformation is $\sqrt{S_{E,1}}$~\cite{zinnjustin2005pi,kleinertbook,altlandsimons}.

To connect two bounces, in principle, one needs to calculate the action with variable ending points.
This is not feasible in practice.
However, since the instanton paths reside mostly near $\phi = 0$, it is justified to expand the action.
Thus, from the initial $\phi(-\beta/2) = \phi_- =0$ to an arbitrary point~$\tilde{\phi}$, the action can be expressed in terms of the final point $\phi(\beta/2) = \phi_+= 0$ as
\begin{align}
    S_{E}[\phi_-, \tilde\phi] \approx S_{E,1}&[\phi_-,\phi_+] \label{eq:action:expansion}\\
    & + \frac{1}{2}\qty(\phi_+ - \tilde\phi)^T \qty[\pdv{S_E}{\phi_i}{\phi_j}] \qty(\phi_+ - \tilde\phi). \notag
\end{align}
The Hessian matrix on the second row is calculated along the classical instanton path.
The same structure is also obtained from a variable initial point and a fixed final point.
Thus, the paths are connected by first dividing the path integral into two parts with a variable mid-point~$\tilde\phi$ and then integrating over it.
These two parts are assumed to obey the instanton solutions individually.

At this point, the multidimensional nature of the problem becomes relevant.
To connect two bounces, I should take into account that the two bounces correspond to different quadratures.
In the case of equal couplings, they are exactly the same. 
Thus, the integration over the mid-point $\tilde\phi$ is a Gaussian integral and the Hessian matrices in Eq.~\eqref{eq:action:expansion} are the same.
In this case, the determinant rising from integration is equal to the inverse of the fluctuation determinant $D_1$~\cite{dashen1974nonperturbative}.
I assume here that this relation holds, at least to an approximation, also in the case of variable coupling constants.

The extension from two to $n$~bounces does not require considerably more effort.
One should note that there are now $n$ hitting times, which are all free parameters and for which the Faddeev-Popov method gives an extraneous factor of $1/n!$.
Otherwise, the fluctuation factor is similar to Eq.~\eqref{eq:onebounce:fluc} for each bounce.
Thus, the general $n$ bounce contribution to the partition function is
\begin{align}
    I^n e^{S_{E}^n} \propto \sqrt{\frac{1}{D}}\sum_{\qty{k}} \frac{(-i \beta)^n}{n!}\prod_{i=1}^n  \sqrt{S_{E,k_i} \epsilon_{k_i}(\beta/n)} e^{-S_{E,k_i}}. 
\end{align}
The vector $k$ enumerates which quadrature hits the wall in each bounce and the sum is taken over all the possible $n$-bounce configurations ($k_i \in \qty{1, \dots N}$).
There are $n$ independent sums and, thus, in total $N^n$ configurations.
These sums can be alternatively written as
\begin{align}
    I^n e^{S_{E}^n} \propto \frac{(-i \beta)^n}{n!} N^n\expval{\sqrt{S_{E} \epsilon(\beta/n)} e^{-S_{E}}}^n. 
    \label{eq:nbounce:contribution:mean}
\end{align}
Here, $\expval{\cdot}$ denotes the ensemble average over the coupling constants~$\lambda^2_i$.

The remaining problem is to calculate the finite temperature correction~$\epsilon(\beta)$ and evaluate the sum over all bounces to arrive at the partition function~$\z$.
The strategy I use is to approximate $\epsilon(\beta/n)^n \approx C(\beta)\epsilon(0)^n$ with a prefactor $C(\beta)$.
This approximation renders the partition function~$\z$ to an exponential form which gives the leading order contribution in temperature to the tunneling rate.
The function $C(\beta)$ plays no role in the rate as it becomes a real prefactor of the imaginary part in the partition function~$\z$.
This approximation is further discussed in Appendix~\ref{app:ftc}.
Effectively, it leads in Eq.~\eqref{eq:nbounce:contribution:mean} to
\begin{align}
    \epsilon_1(\beta/n) \rightarrow \epsilon_1(0) = 4 \omega_{A,1} \omega_{H,1}.
\end{align}
where I need to define the weighted arithmetic average
\begin{align}
    \omega_{A,1} = \frac{(N\expval{\lambda^4} - \lambda_1^4)\omega_0 + \lambda_1^4\qty(\frac{1+\delta}{2}\omega_+ + \frac{1-\delta}{2}\omega_-)}{N\expval{\lambda^4}}
    \label{eq:arithmeticfreq}
\end{align}
with the same weights as in the harmonic average~$\omega_{H,1}$.
Whenever the total coupling $N\expval{\lambda^4}$ is small compared to $\omega_c^2\omega_0^2$ (i.e., the rotating wave approximation is applicable), always $\omega_{A,1} \approx \omega_0$.  
In the limit $\lambda_1^2 \rightarrow 0$, one finds $\epsilon_1(\beta/n)^n = (2\omega_0)^{2n}$ without any approximations.

Finally, the sum over all classical solutions and their quadratic fluctuations can be evaluated to arrive at the partition function~$\z$.
The important quantity here is the average modification~$r$ of the tunneling rate, defined as the ratio of the total tunneling rates with and without the coupling to the cavity, $r = k/k(\lambda=0)$.
I find
\begin{align}
    r = \expval{\frac{\omega_H}{\omega_0}\sqrt{\frac{\omega_A}{\omega_0}} \exp[- S_0
    \qty( \frac{\omega_H}{\omega_0} - 1)]}.
    \label{eq:ratemod}
\end{align}
This analytical result is for an arbitrary distribution of couplings.
It directly shows that the most important polaritonic effects are contained in the harmonic frequency~$\omega_H$ defined in Eq.~\eqref{eq:hf} while the arithmetic mean frequency~$\omega_A$ of Eq.~\eqref{eq:arithmeticfreq} provides a small correction relevant only in the ultra-strong coupling regime.

The rate modification~$r$ describes the total tunneling rate modification of an $N$-body polaritonic system.
The light-matter coupling modifies the tunneling for each system and, thus, there must be an ensemble average over the coupling constants.
To be more precise, the average is over the second-order couplings~$\lambda_i^4$ which are the weighing factors in the harmonic average~$\omega_H$.
Using the expression $\lambda_i^2 = \sqrt{\omega_c \omega_0} g_i$ it is instructive to write
\begin{align}
    \frac{\omega_H}{\omega_0} = \frac{1}{1 
    + \frac{g^2}{N\expval{g^2}}\qty(\frac{1+\delta}{2}\frac{\omega_0}{\omega_+} + \frac{1-\delta}{2}\frac{\omega_0}{\omega_-} - 1)}
\end{align}
in terms of the true coupling constants $g$.
Thus, the relevant distribution is that of $g^2$.
This is in contrast to our recent work focusing on bistable potentials in the thermal activation regime where we found that the distribution of $g$ plays an important role~\cite{kansanen2022cavity}.

\section{Analysis of the polaritonic rate modification}
Let us consider the consequences of the rate modification~\eqref{eq:ratemod}.
In the following, I assume that the rotating wave approximation holds and that $\omega_c \approx \omega_0$.
The polariton frequencies are effectively redefined as
\begin{align}
    \omega_\pm = \frac{\omega_c + \omega_0}{2} \pm \sqrt{N\expval{g^2} + (\omega_c - \omega_0)^2/4}
\end{align}
and $\omega_A/\omega_0 = 1$.
The harmonic average simplifies as the relation
\begin{align}
    \frac{1+\delta}{2}\frac{1}{\omega_+} + \frac{1-\delta}{2}\frac{1}{\omega_-}
    = \frac{\omega_c}{\omega_0\omega_c - N \expval{g^2}}
\end{align}
removes the need for the detuning parameter $\delta$.

\begin{figure}
    \centering
    \includegraphics{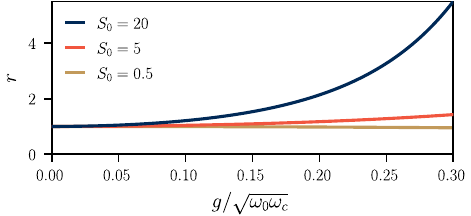}
    \caption{Polaritonic tunneling rate modifications of a single system with different tunneling barriers~$E_b/\omega_0 = S_0/2$.}
    \label{fig:rtm1}
\end{figure}

$N = 1$:
For a single metastable system, the analysis is straightforward.
The harmonic average is then over the polariton states which favors the lower polariton state.
By employing the rotating wave approximation, I have $\omega_H/\omega_0 = 1 - g^2/\omega_0 \omega_c$.
Inserting this relation to Eq.~\eqref{eq:ratemod} gives
\begin{align}
    r = \qty(1 - \frac{g^2}{\omega_c\omega_0})\exp(S_0\frac{g^2}{\omega_c\omega_0}).
    \label{eq:ratemodN1}
\end{align}
Whether the tunneling rate is increased or decreased depends on the bare action $S_0 = 2E_b/\omega_0$.
Expanding to the lowest order in the coupling gives $r \approx 1 + (S_0 - 1)\frac{g^2}{\omega_c\omega_0}$.
A high tunneling barrier is represented by $S_0 > 1$ in which case the rate always increases due to the presence of the cavity.
Higher the barrier, stronger the effect for a fixed coupling~$g$.
This is visualized in Fig.~\ref{fig:rtm1}.
It should be noted that $S_0 < 1$ is at odds with the semiclassical approximation and, thus, the result might not be accurate in such case.

The case of a single tunneling system coupled to a harmonic oscillator is relevant for experiments conducted in superconducting circuits~\cite{ankerhold2007quantum}. 
The metastable quadrature could be, for instance, the superconducting phase difference of a Josephson junction in an electrical circuit.
Then, Eq.~\eqref{eq:ratemodN1} predicts the tunneling rate change if this circuit is connected to an external resonator.

$N \gg 1$ but $N \expval{g^2} < \omega_0 \omega_c$:
The case of macroscopically large $N$ is the typical regime of polaritonic chemistry.
The ensemble average in Eq.~\eqref{eq:ratemod} could be calculated numerically for some model distribution of couplings but, rather, I calculate it with the cumulant expansion to the second order.
This gives
\begin{align}
    r &\approx \qty[\expval{\frac{\omega_H}{\omega_0}} - S_0 \var{\frac{\omega_H}{\omega_0}}] \\
    &\qquad\times
    \exp[S_0 - S_0 \expval{\frac{\omega_H}{\omega_0}} + \frac{1}{2}S_0^2 \var{\frac{\omega_H}{\omega_0}}], \notag
\end{align}
where $\var{\cdot}$ refers to ensemble variance defined as $\var{x} = \expval{x^2} - \expval{x}^2$.
Thus, in principle, the variance of the coupling constants can modify the observed rate modification.
However, for $N \gg 1$, the variance is well-approximated by
\begin{align}
    \var{\frac{\omega_H}{\omega_0}} &=\expval{\frac{\omega_H}{\omega_0}}^4  \frac{\var{g^2}}{(\omega_0 \omega_c - N\expval{g^2})^2},
\end{align}
because the fluctuation of couplings is also suppressed by the factor $1/N$ in the harmonic mean.
Now, if $g_i^2/\omega_c\omega_0 \ll 1$ for all $i$, which is a typical assumption in the collective coupling regime, the variance can be neglected as $\var{g^2/\omega_c\omega_0} \ll 1$.
Consequently, the expectation value of $\omega_H/\omega_0$ is given by
\begin{align}
    \expval{\frac{\omega_H}{\omega_0}} &= \frac{\omega_0 \omega_c - N \expval{g^2}}{\omega_0 \omega_c - (N - 1)\expval{g^2}} \approx 1 - \frac{1}{N}\frac{N\expval{g^2}}{\omega_c\omega_0}. \label{eq:hf:mean}
\end{align}
Here, it appears that $\omega_H/\omega_0$ is determined as a ratio of polariton frequencies~$(\omega_+ \omega_-)^2$ so that the polaritons in the denominator consist of $N-1$~systems and in the nominator of $N$~systems.
The latter equation is an expansion in the leading order of $\expval{g^2}/\omega_0\omega_c$.
Using this expanded form I find
\begin{align}
    r \approx \qty(1 - \frac{\expval{g^2}}{\omega_c\omega_0})\exp(S_0\frac{\expval{g^2}}{\omega_c\omega_0}),
    \label{eq:ratemodN}
\end{align}
which generalizes the single-system polaritonic rate modification of Eq.~\eqref{eq:ratemodN1}.
In conclusion, there is no considerable collective tunneling effect, even if the collective coupling $\sqrt{N\expval{g^2}}$ is a considerable fraction of $\sqrt{\omega_c \omega_0}$.


\subsection{Comparison to high-temperature escape rate}
Thermal activation is the main mechanism in the escape from a metastable potential whenever the temperature is above a threshold temperature proportional to~$\omega_0$~\cite{affleck1981quantum,hanggi1990reaction}.
The instanton path shrinks to a single point in the limit of high temperature, $\beta \rightarrow 0$.
This follows from Matsubara frequency $\omega_{m\neq0} \rightarrow \infty$.
Thus, only $m = 0$ contributes in the Fourier series expressions [e.g. Eg.~\eqref{eq:action:result}].
The action is in this case
\begin{align}
        S_{E,i} = \beta E_b \frac{\omega_0 \omega_c - N \expval{g^2}}{\omega_0 \omega_c - (N\expval{g^2}- g_i^2)}.
\end{align}
The similarity to the low-temperature action in Eq.~\eqref{eq:action:result} is evident: the bare action has changed from $S_0 = 2E_b/\omega_0$ to $\beta E_b$ while the polaritonic modification is expressed in a form similar to Eq.~\eqref{eq:hf:mean} instead of $\omega_{H,i}/\omega_0$.
However, I have not used the rotating wave approximation here as in Eq.~\eqref{eq:hf:mean}.

In the high-temperature regime, one can also calculate the rate using the classical transition state theory~\cite{hanggi1990reaction}.
This approach gives the same action but it allows for a straightforward solution of the factor
in front of the exponent containing the action (also called the attempt frequency).
The rate modification~$r$ obtained in this way is
\begin{align}
    &r = \\
    &\expval{\sqrt{\frac{\omega_0\omega_c - N \expval{g^2}}{\omega_0\omega_c - (N \expval{g^2} - g^2)}}\exp[ \frac{-\beta E_b \cdot g^2}{\omega_0\omega_c - (N \expval{g^2} - g^2)}]}. \notag
\end{align}
The structure of the classical escape rate modification is therefore different from the low-temperature one 
in Eq.~\eqref{eq:ratemod}.
Besides the change of the harmonic frequency~$\omega_H$ to the ratio of polariton frequencies (which coincide in the rotating wave approximation), the modification of the action and pre-exponential factor are in different powers.

With both the low- and high-temperature limits of the rate modification at hand, one can imagine the following set of experiments (see e.g. Ref.~\onlinecite{voss1981macroscopic}):
One varies the temperature of the polaritonic system and measures the escape rate.
Starting from a high temperature and lowering it, the rate drops and eventually saturates to the quantum tunneling rate.
By repeating this measurement without the cavity, the polaritonic coherence effect should become visible.
The results I obtained imply, however, that this is likely only in single systems with sizable light--matter coupling because there is no collective enhancement of the rates.

\section{Conclusion}
The work presented here is rather technical and, in many ways, cumbersome.
Next, I try to clarify what I think are the main ideas and results of the work.

I show a simple, analytically solvable, toy model for polaritonic tunneling.
In principle, there are numerous calculation techniques in the literature but the multidimensionality of the polaritonic system and the ski-jumping potential require some adaptation.
These techniques might prove useful, for instance, in the investigations of macroscopic tunneling in superconducting circuit arrays or other interacting ensembles of metastable systems.

Even if the main result, the polaritonic tunneling rate modification~\eqref{eq:ratemod}, is obtained in a ski-jumping potential that does not directly correspond to any potential seen in nature, it has value.
As a first guess, the structure of the solution is likely similar for a different potential: The modification is determined by the bare action and the harmonic frequency~$\omega_H$.
The formation of polaritons affects the coherence properties in such a way that the tunneling rate may be increased. 
At the same time, if $N \gg 1$, the dark states spoil the effect of the polaritons to the tunneling decay rate out of any metastable potential.
Of course, I would prefer to be proven wrong.

My work in the low-temperature regime coupled with the transition state results in Refs.~\onlinecite{zhdanov2020vacuum} and~\onlinecite{campos2020polaritonic} indicate that there is no collective and resonant polaritonic effect in the escape rate in the case of a large number $N$ of molecules.
However, there are extensions to the model presented here that may affect the end result.
It has been theoretically suggested that inhomogeneous broadening of metastable system frequencies might change polaritonic dark states by providing them some photonic weight which can change the tunneling rate as well~\cite{du2022catalysis,dubail2022large}.
One may also expect changes to tunneling rates when entering the ultrastrong-coupling regime~\cite{forn2019ultrastrong}.
Furthermore, I include a single cavity mode, coupling linearly to a single coordinate of the metastable system.
The inclusion of multiple modes allows for a more realistic description of a cavity and light-matter interaction.
It would open an avenue to investigate possible symmetry effects within light-matter coupling and tunneling rates, motivated by a recent experiment in polaritonic chemistry~\cite{pang2020symmetry}.

This article considers only a truly metastable potential.
An alternative system would be a bistable potential which we have considered in the thermal activation limit~\cite{kansanen2022cavity}.
For the low-temperature limit, the approach would have to be different than what I present here because there are no similar instantons.
This is because these imaginary-time paths are at zero energy while the cavity changes the energies of the stationary states. 
Tunneling in bistable systems therefore requires another approach.

I did not take into account the friction or dissipation the systems realistically have.
On the level of the action this would be, in principle, a straightforward extension~\cite{caldeira1981influence,weiss2012quantum,altlandsimons,ankerhold2007quantum}.
I expect dissipation to modify the tunneling rate modification: 
since the formation of polaritons leads to a coherent effect, the modification should be larger for a nearly dissipationless cavity than for a bad cavity with a large dissipation rate.
However, it should not change, for instance, the $N$-scaling of the action.

\begin{acknowledgements}
I thank Tero Heikkilä for useful discussions and for coining the term ``ski-jumping potential''.
This work has been supported by the Magnus Ehrnrooth foundation and the Academy of Finland (project numbers 317118 and 321982).
\end{acknowledgements}

\appendix

\section{An approach to classical multi-bounce solutions}\label{app:multibounce}
For completeness, I show how the method presented in the main text can be expanded to finding classical solutions
with multiple quadratures visiting the point $q_i = a$ at specified times $\tau_i$.
Even though such instanton configurations do not contribute to the quantum tunneling rate (see main text), the classical problem is interesting on its own.
The approach presented in the main text requires modification.

One has to introduce a matrix-like structure to the delta functions in the equations of motion.
From the viewpoint of Lagrangian mechanics, the delta function can be seen as a constraint force following from the condition $q_1(\tau_1) = a$.
If there is another constraint, say, $q_2(\tau_2) = a$, the constraint on quadrature~1 affects also quadrature~2 due to the coupling via the cavity.
The equations of motion to be solved are in general
\begin{align}
    - \ddot{q}_i + \omega_0^2 q_i + \lambda_i^2 x &= \sum_j A_{ij} \delta(\tau - \tau_j)
\end{align}
for all quadratures that hit the wall, $q_i(\tau_i) = a$.

If one assumes that all the off-diagonal elements of $A_{ij}$ are zero, the delta functions do not conserve energy~$E(\tau) = \frac{1}{2}\dot{\phi}^T \dot{\phi} - V_\mathrm{tot}(\phi)$.
(The energy is zero for instanton solutions in the limit $\beta \rightarrow \infty$.)
That is, the solution is only correct in a piecewise manner with abrupt changes in energy at hitting times~$\tau_i$.
The energy is conserved for a single bounce.

One can fix the unknown parameters $A_{ij}$ not only from the conditions $q_i(\tau_i) = a$ but also from the energy conservation and symmetry considerations.
This seems clear for two bouncing quadratures, because the constraint forces between the two systems should be similar and, consequently, $A_{12} = A_{21}$. 
Whether or not it can be adapted the solution of the general many-bounce dynamics, I do not know.

When there is a solution of the action in terms of the hitting times $\tau_i$ which are fixed in the beginning, the exact meaning of the solution is still unclear in the viewpoint of classical mechanics.
Presumably, in the spirit of the least action, the classical path should be that of minimal action.
Thus, one should further minimize the action in terms of the hitting times~$\tau_i$ which presents another difficult step in a general case.

Finally, I note that in the special case $\tau_i = 0$ and $\lambda_i^2 = \lambda^2$ for $i = 1,2,\dots M$ there is a much more straightforward route to the solution.
Then, the $N$ quadratures can be divided into two classes: those that hit the wall and those that do not.
The equations of motion are exactly the same within these two classes.
Therefore, the dynamics can be described by using $Q_1 = \sum_{i=1}^M q_i$ and $Q_0 = \sum_{i=M+1}^N q_i$ for which the equations of motion in the inverted potential are
\begin{subequations}\label{eq:collectiveEOMS}
\begin{align}
    - \ddot{x} + \omega_c^2 x +  \lambda^2 (Q_0 + Q_1) &= 0, \\
    - \ddot{Q}_1 + \omega_0^2 Q_1 + M\lambda^2 x &= M A \delta(\tau),\\
    - \ddot{Q}_0 + \omega_0^2 Q_0 + (N-M)\lambda^2 x &= 0.
\end{align}
\end{subequations}
This set of equations can be solved in a similar way as the one-bounce problem in the main text [note that $Q_1(0) = M a$].
This results in the action
\begin{align}
    S_{E,M} = \frac{M}{2} a^2 \qty[\frac{1}{\beta}\sum_m \frac{1 + M\lambda^4 \chi_P(\omega_m)}{\omega_0^2 + \omega_m^2}]^{-1},
\end{align}
where $\chi_P(\omega_m)$ is as in Eq.~\eqref{eq:polresponse}.
In the limit of no light-matter coupling, $S_{E,M} = M S_{E,1}$ which reads $S_{E,M} = \beta M E_b$ in the high-temperature limit.
This limit shows that the action scales with $M$ for fully independent systems.
The multi-bounce trajectories should be neglected in the tunneling rate calculation: They describe configurations that will not turn out to be saddle points of the action.

\section{Finite temperature correction}\label{app:ftc}
For large but finite $\beta$, an exponentially small correction to the zero eigenvalue mode should be included.
This section generalizes the discussion in Ref.~\onlinecite{gildener1977pseudoparticle} to multidimensional systems.
Since the eigenvalues of the fluctuation determinant can be mapped to the eigenvalues in the time-independent Schrödinger equation, the question is, how do the energies change when a system is put into an infinite potential well.
That is, there are two equations
\begin{subequations}
\begin{align}
    (\partial_\tau^2 + \mathcal{V})f(\tau) &= 0, \label{eq:ftc:fluceq1}\\
    (\partial_\tau^2 + \mathcal{V})g(\tau) &= -\epsilon g(\tau) \label{eq:ftc:fluceq2}
\end{align}
\end{subequations}
with boundary conditions $f(\tau \rightarrow \pm \infty) \rightarrow 0$ and $g(\tau = \pm \beta/2) = 0$.
It can be shown that the solution of the former equation is related to the classical instanton path~$\phi(\tau)$ by $f(\tau) = \dot\phi(\tau)$.
By multiplying Eq.~\eqref{eq:ftc:fluceq1} by $g^T$ from the right and similarly Eq.~\eqref{eq:ftc:fluceq2} by $f^T$,  integrating over $[-\beta/2, \beta/2]$, and then subtracting the equations, I find
\begin{align}
    f^T(-\beta/2)\dot g(-\beta/2) - &f^T(\beta/2)\dot g(\beta/2) \notag\\
    &= \epsilon\int_{-\beta/2}^{\beta/2} f^T(\tau) g(\tau) \dd{\tau}.
    \label{eq:ftc:temp1}
\end{align}
Since the correction must be small for large $\beta$, the integral on the right hand side can be approximated by replacing~$g$ by~$f$ and extending the integration limits to~$\pm \infty$.
Consequently, $\epsilon$ can be solved in terms of $f$ and $g$.
The question is then about the relation between the derivative of $g$ and $f$ at the boundaries~$\pm \beta/2$.

For real-valued functions, the WKB approximation gives $\dot g(\pm \beta/2) \approx 2 \dot f(\pm\beta/2)$.
Alternatively, one can set $g(\tau) = c(\tau) f(\tau)$ and find $c$ to first order in $\epsilon$ which results in 
$\dot g(\pm \beta/2) \approx 2\sqrt{3} \dot f(\pm\beta/2)$ for the ski-jumping potential.
However, in this article, the exact proportionality constant is not of great importance as such factors cancel out when determining the cavity-induced modifications to the tunneling rate.

For vector-valued $f$ and $g$, the argument is similar but less rigorous.
It is possible that putting the system into a box changes both the magnitude of the derivative and its direction.
However, the change of direction can be neglected in the ski-jumping potential:
The potential matrix $\mathcal{V}$ is close to a constant near the boundaries so one can diagonalize it by an orthogonal matrix.
For the ski-jumpinging potential especially, $\mathcal{V}$ is constant for all values of~$\tau$ except the hitting time $\tau = \tau_1$.
The argument for real-valued functions holds then for each component of the transformed vectors.
Since the transformation is the same for both $f$ and $g$, the result is also the same.
Inserting the relations $\dot g(\pm \beta/2) \approx 2 \dot f(\pm\beta/2)$ and $f = \dot \phi$ into Eq.~\eqref{eq:ftc:temp1}, I get Eq.~\eqref{eq:finiteTcorr}.
It should be noted, however, that this result will likely not hold for more complicated potentials but, in general, $\dot g(\pm\beta/2) = C \dot f(\pm\beta/2)$ where $C$ is a matrix.

\subsection{Correction for ski-jumping potential}

\begin{figure}
    \centering
    \includegraphics{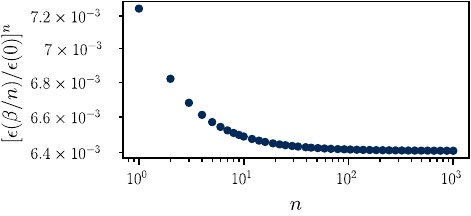}
    \caption{The values of $[\epsilon(\beta/n)/\epsilon(0)]^n$ for $N = 1$, $\beta \omega = 5$ and $(1 \pm \delta)\omega_\pm/\omega_0 = 1 \pm 0.1$.}
    \label{fig:bounds}
\end{figure}

The finite temperature correction~\eqref{eq:finiteTcorr} is readily obtained by using the instanton solutions~\eqref{eq:instpath}.
It should be noted that the denominator in the correction~$\epsilon_1(\beta)$ is the action~$S_{E,1}$ and that the instanton solutions are symmetric with respect to the hitting time~$\tau_1=0$.
I find
\begin{align}
    \epsilon_1(\beta) = 4 &\omega_{H,1} \frac{1}{N\expval{\lambda^4}} \Bigg[(N\expval{\lambda^4} - \lambda_1^4)\omega_0 e^{-\beta \omega_0} \\
    &+ \lambda_1^4\qty(\frac{1+\delta}{2}\omega_+ e^{-\beta \omega_+} + \frac{1-\delta}{2}\omega_-e^{-\beta \omega_-}) \Bigg]. \notag
\end{align}
As implied in the main text, this expression is not particularly helpful because the partition function~$\z$ depends on $\epsilon(\beta/n)$ so that $n$ is summed over.
A reasonable approximation is to replace $\epsilon(\beta/n)$ by $\epsilon(0)$ because it represents the $n\rightarrow \infty$ limit.
Furthermore, it can be shown that, for a constant~$\beta$, the value of $\epsilon(\beta/n)$ can be limited by $\epsilon(0)$ in the sense that $A \leq [\epsilon(\beta/n)/\epsilon(0)]^n \leq B$ for all~$n$ with suitable constants~$A$ and~$B$.
This is exemplified in Fig.~\ref{fig:bounds}.
The prefactor in the approximation $\epsilon(\beta/n) \propto \epsilon(0)$ does not contribute to the imaginary part of the partition function~$\z$ and is thus unimportant.


\begin{thebibliography}{40}
\bibitem [{tst()}]{tstnote}%
  \BibitemOpen
  \href@noop {} {}\bibinfo {note} {More technically, the Hessian
  matrix~$\mathcal{H}$ with elements~$\mathcal{H}_{ij} =
  \pdv*{V_\mathrm{tot}}{\phi_i}{\phi_j}$ evaluated at a point where
  $\partial_{\phi_i}V_\mathrm{tot} = 0$ for~$i=1,2,\dots N$ must have 1
  negative eigenvalue and $N-1$ positive eigenvalues.}\BibitemShut {Stop}%
\end{thebibliography}
\end{document}